\documentclass[%
 reprint,
nofootinbib,
 amsmath,amssymb,
 aps,
]{revtex4-2}

\usepackage{graphicx}% Include figure files
\usepackage{dcolumn}% Align table columns on decimal point
\usepackage{bm}% bold math
\begin{document}

\preprint{APS/123-QED}

\title{A survey of the electroweak configuration space and the $W$ boson mass}% Force line breaks with \\

\author{Oswaldo Vazquez}
 \email{ovazquez at college.harvard.edu}
\affiliation{%
 Department of Mathematics, Harvard University, 1 Oxford Street, MA 02138, USA}%

\begin{abstract}
Following the recent work of V. Moncrief, A. Marini, R. Maitra \cite{moncrief} and P. Mondal \cite{mondal1} on the geometry of field theoretic configuration spaces, this account examines how the regularized Ricci curvature of the $SU(2)_L \times U(1)_Y$ Yang-Mills orbit space may provide an intrinsic mass to the $W$ boson which contributes to the value obtained from the renormalized Higgs mechanism. Though the discussion is heuristic, one hopes that this infinite-dimensional technology, which does not postulate extensions to the Standard Model, could explain the mass anomaly reported by the CDF II collaboration.
\end{abstract}

%\keywords{Suggested keywords}%Use showkeys class option if keyword
                              %display desired
\maketitle

%\tableofcontents

\section{\label{sec:1}Background and geometric motivations}

The direct measurement of the $W$ boson mass by the CDF II detector \cite{cdf} has caught many eyes due to its significant deviation from the Standard Model (SM) prediction \cite{pdg} and results from the ATLAS collaboration \cite{atlas}. In turn, several theoretical proposals have been made to reconcile the discrepancy albeit by introducing extensions to the SM such as a $U(1)$ dark sector \cite{dark}, a Higgs triplet \cite{higgstriplet}, minimal supersymmetry \cite{mss} and more \cite{FileviezPerez:2022lxp,Krasnikov:2022xsi,Nagao:2022oin,Lee:2022nqz,Bahl:2022xzi,Kawamura:2022uft,Yuan:2022cpw,heckman}. The call to probe for new physics is strong and exciting, yet it is worth asking whether the apparent anomaly can be explained without any extra ingredients. It is the goal of this presentation to argue for a positive answer to said question by exploring the \textit{true} origin of mass in gauge theories.

As known, the Higgs mechanism of spontaneous symmetry breaking (SSB) is responsible for endowing the weak bosons with masses at the classical level while keeping renormalizability \cite{masslessrenorm,massiverenorm} and gauge invariance of a reduced structure group. However, evidence from perturbation theory in the context of quantum chromodynamics (most notably the phenomenon of asymptotic freedom \cite{asymfree}) seems to point in the direction that quantum Yang-Mills theories with compact semi-simple non-Abelian structure groups over $\mathbb{R}^{1,3}$ possess a mass gap despite the lack of an explicit mass term in the action pre-SSB. A crude but somewhat supportive case for a mass gap goes as follows. Perturbatively, one breaks the gauge symmetry by splitting the action into a solvable part (consisting of several copies of $U(1)$ gauge fields) and then treat the other terms as interactions. According to the Källén–Lehmann representation theorem, the lowest pole in the two-point function for the full theory should give us information about the mass-squared of the first excited state of the Hamiltonian. Let $1/p^2$ roughly denote the free $U(1)$ propagator, then one is unable to yield a mass at any finite order of perturbation since the pole is zero
\begin{equation}
    \frac{1}{p^2}+\frac{1}{p^4}+\frac{1}{p^6}+\cdots+\frac{1}{p^{2k}}, \quad \mathrm{pole=0}\nonumber
\end{equation}

Keeping in mind issues of convergence, adding all possible terms in the schematic expansion for the 2-point function does give a mass as the pole becomes nonzero
\begin{equation}
    \frac{1}{p^2}+\frac{1}{p^4}+\frac{1}{p^6}+\cdots= \frac{1}{p^2-1}, \quad \mathrm{pole=1}\nonumber
\end{equation}

These ideas are of course folly compared to the rigorous demands of constructive quantum field theory \cite{qym}. Nevertheless, if existence of quantum Yang-Mills with mass gap proves to be true then it might be possible to give rise to intrinsic masses for the weak bosons in addition to those obtained from the Higgs mechanism. In particular, it is desirable for the intrinsic mass of the $W$ to be small enough to account for the CDF II discrepancy. In no way does this presentation intend to tackle the monumental problem of existence of quantum Yang-Mills, rather the result is assumed in order to study the source of mass gap in the electroweak theory by following the analysis of V. Moncrief, A. Marini, R. Maitra \cite{moncrief} and P. Mondal \cite{mondal1}.

Begin by recalling the Yang-Mills action on $\mathbb{R}^{1,3}$, 
\begin{equation}
\label{eq:action}
    I[A]=-\frac{1}{4}\int_{\mathbb{R}^{1,3}} ||F_{A}||^2
\end{equation}
where $A$ is a connection of a $G$-bundle over Minkowski space, $F_A=dA+g[A,A]$ is the associated curvature 2-form with $g$ denoting the coupling, and $||\cdot||^2=\langle \cdot, \cdot \rangle$ is a positive-definite adjoint-invariant inner product on the Lie algebra $\mathfrak{g}$ (this is possible since $G$ is assumed to be a compact matrix group). If two connections $A_1$ and $A_2$ are defined over the same local trivializing set of $\mathbb{R}^{1,3}$, then acting by a transition function (local gauge transformation) $U(x)\in G$ yields gauge invariance of the theory $A_2(x)\sim U(x)^{-1}A_1(x)U(x)+U(x)dU(x)^{-1}$. The dynamics of a solution to the field equations is described by a regular curve $t\mapsto (A(t),\pi_A(t))$ with prescribed initial conditions living on the phase space. It is important to note that the affine space of spatial connections $\mathcal{A}$ is \textit{not} the honest configuration space due to the aforementioned gauge invariance property, instead one must quotient $\mathcal{A}$ by the group of reduced gauge transformations $\mathcal{G}$\footnote{The group $\mathcal{G}$ includes ordinary gauge transformations as well as those that stabilize a connection i.e. $A=U^{-1}AU+UdU^{-1}$. This gets rid of residual gauge freedom.}. Thus, the classical physics takes place in $\mathcal{A}/\mathcal{G}$. The nature of this orbit space has been studied throughout the decades \cite{singer,orbittopology,orbitgeometry}, it is an infinite-dimensional manifold with incredibly non-trivial topology and moreover it comes equipped with a Riemannian metric $\mathfrak{G}$ induced from the kinetic part of $I[A]$. If we denote coordinates on the base by $x^\mu=(t,\vec{x})^\mu$ and the Lie algebra generators by $T^{a}$, then the metric takes the following form on the local Coulomb chart around $A=A_{\mu}^{a}T^{a}dx^{\mu}=0$ (equivalently, the gauge condition $\partial_i A_i^a=0$)
\begin{eqnarray}
\label{eq:metric_orbit}
    \mathfrak{G}[A]_{{A^{a}_i(\vec{x})}{A^{b}_j(\vec{y})}}&=&\delta_{ij}\delta^{ab}\delta(\vec{x}-\vec{y}) \nonumber \\
    &&+f^{eca}A_{i}^{c}(\vec{x})\Delta_A^{-1}(\vec{x},\vec{y})f^{edb}A_{j}^{d}(\vec{y})
\end{eqnarray}
with $f^{abc}$ being the structure constants with respect to the chosen basis for $\mathfrak{g}$ and $\Delta_A=(D_{A})^2=(\partial+g[A,\cdot])^2$ is the gauge covariant Laplacian. The metric is manifest at the level of the action functional as
\begin{eqnarray}
    I[A]&=&\int dtd\vec{x}d\vec{y}\;\frac{1}{2}\mathfrak{G}[A]_{{A^{a}_i(t,\vec{x})}{A^{b}_j(t,\vec{y})}}\dot{A}_{i}^{a}(t,\vec{x})\dot{A}_{j}^{b}(t,\vec{y}) \nonumber \\
    &&-\frac{1}{4}\int dtd\vec{x}\; F^{a}_{ij}(t,\vec{x})F^{a}_{ij}(t,\vec{x})\nonumber
\end{eqnarray}

\begin{figure}[b]
\includegraphics[scale=0.3]{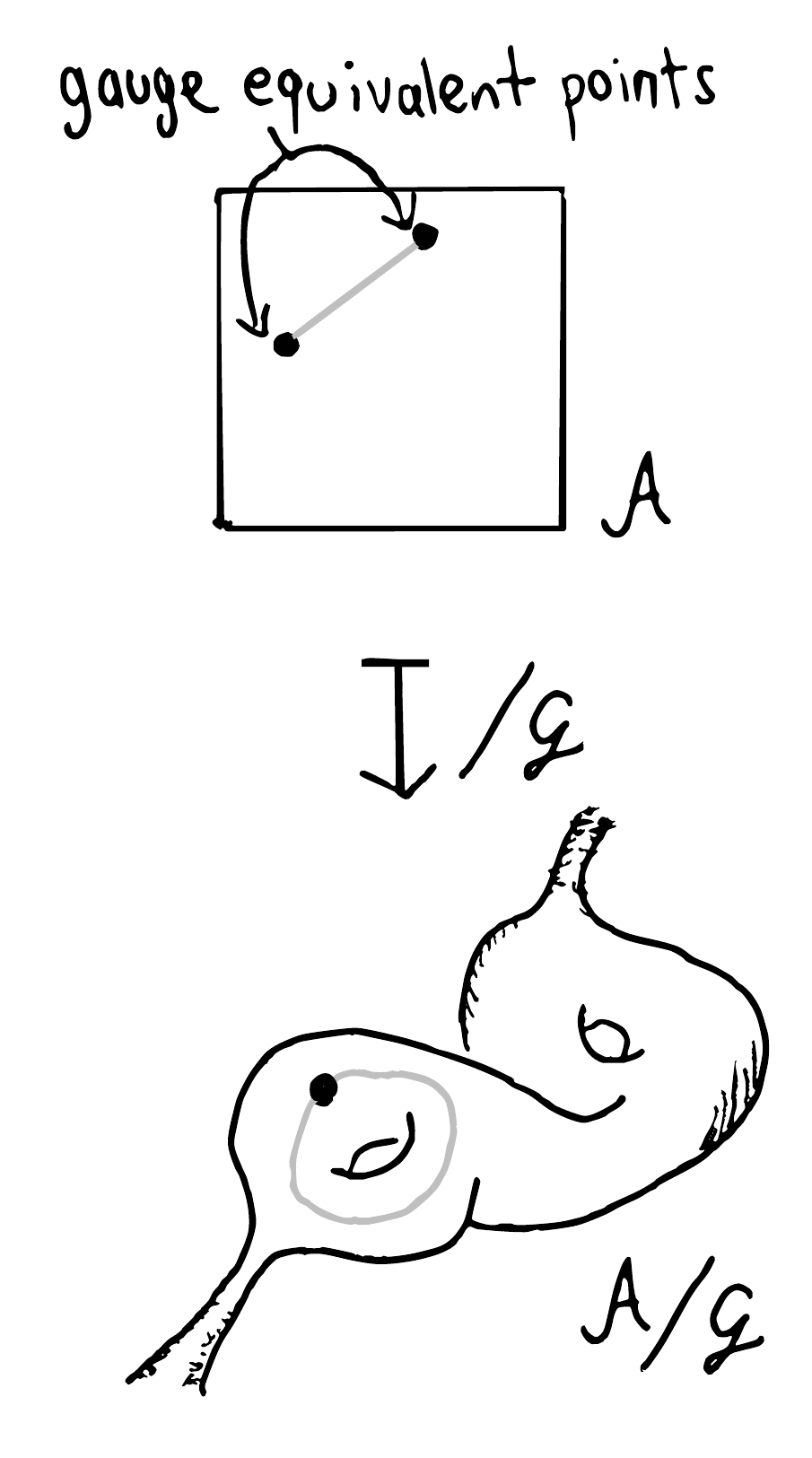}% Here is how to import EPS art
\caption{\label{fig:orbit_space} The flat affine space $\mathcal{A}$ becomes curved upon taking the quotient with $\mathcal{G}$.}
\end{figure}

Note that explicit calculations for the above can be found in \cite{mondal1}. From (\ref{eq:metric_orbit}) one can see that $\mathcal{A}/\mathcal{G}$ is curved (Fig. \ref{fig:orbit_space}), therefore the canonical quantization procedure will identify the conjugate momentum $\pi_A$ with $-i\hbar$ times the Levi-Civita covariant derivative on the orbit space $\mathfrak{D}_{A}=\delta_{A}+\Gamma$ (the $\delta_{A}$ represents the functional derivative with respect to $A$). The Hamiltonian is then a formal Laplace-Beltrami operator-valued distribution on $\mathcal{A}/\mathcal{G}$ plus the potential term
\begin{equation}
\label{eq:hamiltonian}
    H=-\frac{\hbar^2}{2}\mathfrak{D}_{A}^2+\frac{1}{4}\int_{\mathbb{R}^3}F^{a}_{ij}F^{a}_{ij}
\end{equation}

Estimating the spectral gap of $H$ becomes tractable as long as one makes suitable assumptions about the quantum Yang-Mills theory, specifically the existence of a normalizable ground state wave functional and a well-behaved weighted measure on the orbit space. This is the content of the work from P. Mondal \cite{mondal1} which was hugely inspired by the intuition of R.P. Feynman who thought that the curvature of the orbit space and the effects of the non-trivial potential might have a bearing on the mass gap \cite{feynman}.

Feynman's idea was to see the extent to which the gap obtained in 1D particle in a box generalizes to the case of infinite-dimensional geometry. In 1D, the ground state is nodeless and real-positive, which means that the first excited state must contain positive and negative contributions due to orthogonality. This defines a certain ``distance" $\mathfrak{L}$ between the positive and negative parts which cannot be arbitrarily large since the potential well forces the quantum mechanical problem to be defined on a compact region of $\mathbb{R}$. Because the particle is free inside the box, the excited energy is purely kinetic and must scale as $1/\mathfrak{L}^2$ due to the Laplacian. Transitioning to the case of our infinite-dimensional orbit space, the ground state wave functional $\Psi_0[A]$ is presumably nodeless and real-positive due to the potential term being independent of time derivatives of configurations. Taking the ansatz $\Psi_0[A]=N_\hbar e^{-S[A]/\hbar}$ where $S[A]$ is real-valued and $N_\hbar$ is a proportionality constant, the normalization condition reads
\begin{equation}
\label{eq:ground_norm}
    \int_{\mathcal{A}/\mathcal{G}}|N_\hbar|^2e^{-2S[A]/\hbar}\sqrt{|\mathrm{det}(\mathfrak{G})|}=1
\end{equation}

\begin{figure*}
\includegraphics[scale=0.8]{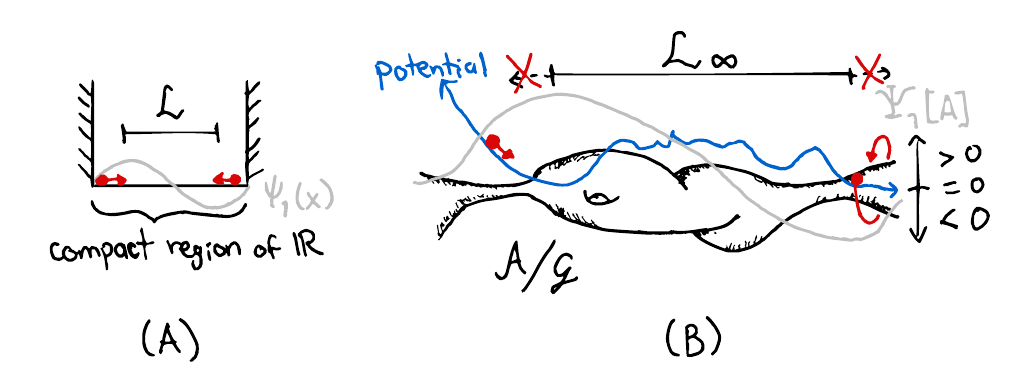}% Here is how to import EPS art
\caption{\label{fig:feynman}(A) represents the 1D particle in a box, the distance $\mathfrak{L}$ is not arbitrarily large due to compactness of the configuration space, point particles (in red) cannot move past the potential well. (B) is a visual of the qualitative behavior that may occur in the orbit space. The potential growth is such that a particle (in red again) cannot escape towards the left noncompact direction. Despite the potential vanishing in the right noncompact direction, the positive curvature of the space helps confine the particle. The distance $\mathfrak{L}_\infty$ between the positive and negative regions of the excited wave functional $\Psi_1[A]$ cannot be extended to infinity due to the aforementioned effects of the geometry \textit{and} the potential.}
\end{figure*}

The potential contains quartic terms in the connection as well as lower order combinations which include spatial derivatives, these terms must of course be well-behaved (in a suitable function space setting) for the potential to rise fast enough and yield confinement of $\Psi_0[A]$. This effect will obviously influence the form of $S[A]$ due to the condition (\ref{eq:ground_norm}) which then supplies a \textit{weighted} measure $\mu(\mathfrak{G},S):=e^{-2S[A]/\hbar}\sqrt{|\mathrm{det}(\mathfrak{G})|}$ to $\mathcal{A}/\mathcal{G}$. The first excited state must be $L^2$-orthogonal to $\Psi_0[A]$, meaning that it must contain orbit space regions where it is positive and other ones where it is negative as in 1D. This property again defines a distance $\mathfrak{L}_\infty$ between the positive and negative regions. The question is whether such distance has a finite upper bound, if so then we can be confident in obtaining a gap. However, closed and bounded balls in infinite dimensions fail to be compact, meaning we lose this privilege from the 1D particle in a box. The naive guess is that the distance can be made as large as we want in non-compact directions, but not all hope is lost when we realize that the \textit{weighted} curved nature of the orbit space can aid in restricting the size of $\mathfrak{L}_\infty$. Indeed, the weighted curvature includes the typical curvature from $\mathfrak{G}$ and effects from the potential made manifest in $S[A]$. As discussed, the potential will tend to confine but it may still admit flat directions and it is at that stage that the curvature contribution solely from $\mathfrak{G}$ will need to step in to restrict $\mathfrak{L}_\infty$. In particular, the desired behavior is obtained when the $\mathfrak{G}$-Ricci curvature is positive---this makes intuitive sense at least from a pictorial point of view (see Fig. \ref{fig:feynman}) since positivity entails a sort of ``converging" behavior. On the other hand, whenever $\mathbf{Ric}_{\mathfrak{G}}$ vanishes then we expect $S[A]$ to yield the mass gap.

From the standpoint of geometric analysis, these ideas would hint at a generalization of the Lichnerowicz theorem \cite{Lichnerowicz1,Lichnerowicz2} for estimating the first positive eigenvalue of the Laplace-Beltrami operator on compact finite-dimensional manifolds. In the current context, we ditch compactness and finite dimensions and replace the Laplace-Beltrami operator with (\ref{eq:hamiltonian}). The finite case requires positivity of Ricci while the orbit space would need positivity of a \textit{weighted} Ricci which depends on $\mathfrak{G}$ and $S[A]$. The speculations discussed are confirmed by a recent theorem due to P. Mondal \cite{mondal1}.

\textbf{Theorem:} \textit{Under the assumption of a 1+3 quantized Yang-Mills theory with normalizable ground state} $\Psi_0[A]=N_\hbar e^{-S[A]/\hbar}$, \textit{the renormalized Hamiltonian has a mass gap} $\Delta>0$ with
\begin{equation}
    E^{(1)}-E^{(0)}\geq \frac{\hbar^2}{2}\Delta \nonumber
\end{equation}
\textit{as long as the renormalized Bakry–Émery Ricci curvature} $\mathbf{Ric}^{B-E}:=\mathbf{Ric}_{\mathfrak{G}}+\mathbf{Hess}_S$ \textit{verifies the following lower bound}
\begin{equation}
    \mathbf{Ric}^{B-E}(\alpha,\alpha)\geq\Delta\mathfrak{G}(\alpha,\alpha) \nonumber
\end{equation}
\textit{for any tangent vector $\alpha\in T(\mathcal{A}/\mathcal{G})$.}

A few remarks need to be made about this theorem. First, the notion of renormalized geometry and Hamiltonian is necessary due to the infinite degrees of freedom. In particular, one sees divergence in the metric (\ref{eq:metric_orbit}) as $|\vec{x}-\vec{y}|\rightarrow 0$ and similarly in the $H$ operator because the Laplace-Beltrami on $\mathcal{A}/\mathcal{G}$ is singular already at leading order. \cite{mondal1} introduces a point-split regulator $\chi$ and counterterms to obtain the appropriate finite parts of $H$ and $\mathbf{Ric}^{B-E}$, a notable consequence is the emergence of an energy scale via $\chi$ based on dimensional grounds which remains present after renormalization of $\mathbf{Ric}_{\mathfrak{G}}$ at the flat connection $A=0$
\begin{equation}
    \mathbf{Ric}_{\mathfrak{G}}^{ren}[A=0](\alpha,\alpha)=\frac{3C(\mathfrak{g})\chi g^2}{2\pi^3}\int d\vec{x}d\vec{y}\; \alpha^{a}(\vec{x})\alpha^{a}(\vec{y})\nonumber
\end{equation}
where $C(\mathfrak{g})$ is the Casimir element of the adjoint representation of $\mathfrak{g}$. The above indicates that $g^2$ must depend on $\chi$ to erase any trail of the regulator and yield a finite result. In other words, $g$ becomes a running coupling.

Secondly, Mondal shows via a heat kernel argument that the renormalized Ricci is positive at the flat connection and away from it. Concretely, the inequality $\mathbf{Ric}_{\mathfrak{G}}^{ren}[A\neq0]>\mathbf{Ric}_{\mathfrak{G}}^{ren}[A=0]\geq 0$ is proven. The immediate upshot is that the gap will satisfy $E^{(1)}-E^{(0)}\geq \lim_{\chi\rightarrow\infty}\frac{\hbar^23C(\mathfrak{g})\chi g(\chi)^2}{4\pi^3}+(\mathbf{Hess}_S$ contribution) provided the second term is non-negative so that it does not cancel the strictly positive contribution from the Ricci part. There is yet to be a rigorous proof for the non-negativity of the Hessian contribution due to its immense difficulty (specific obstacles are discussed in sections 3 and 6 of \cite{mondal1}), nevertheless explicit computations for the ground state wave functional can be made in the high and low energy limit \cite{hatfield,lowenergylimit}. In both cases, the Hessian term turns out to be non-negative \cite{mondal2}. Lorentz invariance of the theory would hint at full non-negativity because the potential cannot be independent of the geometrically rich kinetic term, thus the $S[A]$ functional must also be ruled by $\mathfrak{G}$ and in turn produce a term of the form $\sqrt{m_0^2+|\vec{k}|^2}$ ($m_0\geq0$ and $\vec{k}$ a continuous 3-momentum) plus measure zero contributions. This bare-bones claim is motivated by the work of \cite{1+2,1+2ground} in 1+2 dimensions and \cite{1+3} in 1+3. For example, \cite{1+2ground} found the $S[A]$ functional in 1+2 dimensions to be
\begin{equation}
    S[A]=\frac{1}{2g^2}\int_{\mathbb{R}^2\times\mathbb{R}^2} \frac{B^{a}(\vec{x})B^{a}(\vec{y})}{m_0+\sqrt{m_0+\Delta_{\mathbb{R}^2}}}\nonumber
\end{equation}
where $B^{a}$ is the chromo-magnetic field and $\Delta_{\mathbb{R}^2}$ is the Laplacian on $\mathbb{R}^2$. The present discussion will take such claim as true and posit the $m_0$ quantity as the classical mass found at the level of the Lagrangian based on the calculations for $\mathbf{Hess}_S$ in solvable theories like Maxwell electromagnetism and the free massive Klein-Gordon theory (see \cite{mondal2} for reference). For the electroweak model, the Hessian contribution to the total mass of the first excited states will be assumed to come solely from the classical Higgs mechanism mass terms whereas the Ricci part will yield intrinsic mass corrections that are quantum in origin (i.e. they cannot be simply inferred from the renormalized Lagrangian but instead one must study the geometry of the electroweak configuration space).

\section{\label{sec:2}orbit space of the electroweak bosonic sector}

The bosonic sector of the Weinberg-Salam-Glashow electroweak model \cite{Weinberg,Salam,Glashow} consist of a Yang-Mills connection $A$ with gauge group $SU(2)_L\times U(1)_Y$. The weak isospin and weak hypercharge generators are denoted by $\{T^{a}\}_{a=1,2,3}$ and $Y=T^4$, respectively. At the classical level, all four bosonic fields are massless until a postulated Higgs field transforming in a 2D representation of $SU(2)_L\times U(1)_Y$ spontaneously breaks the gauge symmetry by acquiring a nonzero vacuum expectation value. This process retains a smaller $U(1)_{EM}$ symmetry corresponding to electromagnetism which is generated by the hypercharge and the maximal torus of $SU(2)_L$. The Lagrangian then reorganizes itself to yield classical masses to the force carriers that we see in nature
\begin{eqnarray}
    \label{eq:ew_lagrangian}
    \mathcal{L}_{EW}&=&-\frac{1}{4}B_{\mu\nu}B^{\mu\nu}-\frac{1}{4}Z_{\mu\nu}Z^{\mu\nu}-\frac{1}{2}W^+_{\mu\nu}W^{-\mu\nu} \nonumber\\
    &&+\frac{1}{2}m_Z^2Z_{\mu}Z^{\mu}+m_W^2W^+_{\mu}W^{-\mu}\\
    &&+\frac{1}{2}\partial_\mu h\partial^\mu h-\frac{1}{2}m_h^2 h^2+\mathrm{interactions} \nonumber
\end{eqnarray}
where $Z$ and $B$ are obtained from a rotation in the $(A^{3},A^{4})$-plane by a weak mixing angle $\theta_W$ and $W^\pm$ are holomorphic and antiholomorphic coordinates in the complexified $(A^{1},A^{2})$ space. The curvature 2-forms are given by $X_{\mu\nu}=\partial_\mu X_\nu-\partial_\nu X_\mu$ with $X=B,Z,W^\pm$. Under a $U(1)_{EM}$ action generated by $\Lambda$, the physical fields transform as follows
\begin{eqnarray}
    &B\mapsto B+g_{U(1)_Y}^{-1}d\Lambda \nonumber \\
    &Z\mapsto Z \\
    &W^\pm \mapsto W^\pm \pm i\Lambda W^\pm \nonumber
\end{eqnarray}
and they leave the theory invariant. Based on these considerations, one may extract the metric of the electroweak orbit space from the kinetic part of the action. The Abelian gauge symmetry of the photon $B$ implies that the metric will take the form of (\ref{eq:metric_orbit}) but without the second term as the structure constants vanish. Similarly, the fact that $Z$ does not transform at all enforces its orbit space to also be flat 
\begin{eqnarray}
    \mathfrak{G}_{Z_i(\vec{x})Z_j(\vec{y})}=\mathfrak{G}_{B_i(\vec{x})B_j(\vec{y})}=\delta_{ij}\delta(\vec{x}-\vec{y})
\end{eqnarray}
One can immediately conclude the vanishing of the Ricci curvature along the $B$ and $Z$ directions, therefore the total masses for these particles will only be attributed to the Higgs mechanism which is expected to be manifest in the Hessian term of the gap theorem.

A slighter challenge is faced when attempting to compute the metric in the $W^\pm$ directions due to its non-gauge-like transformation, this will correspond to us obtaining a curved space. Start by reconsidering the theory before SSB and focus on the time component of the Yang-Mills equation with Higgs source
\begin{equation}
    D_\mu F^{0\mu}[A]=J^{0}[h,h^\dagger], \quad J^{\mu}[h,h^\dagger]=ihD^{\mu}h^\dagger+h.c.
\end{equation}
Using the Coulomb gauge condition yields the following formal expression for $A_0$
\begin{equation}
\label{eq:a_0}
    A_0=\Delta_A^{-1}([A_i,\dot{A}_i]-J^{0}[h,h^\dagger])
\end{equation}

The $SU(2)_L \times U(1)_Y$ orbit space metric $\mathfrak{G}^{EW}$ is then extracted from the kinetic term of the action as
\begin{eqnarray}
    \mathfrak{G}^{EW}(\dot{A},\dot{A})&=&\int d\vec{x}d\vec{y}\;\mathfrak{G}^{EW}_{{A^{a}_i(\vec{x})}{A^{b}_j(\vec{y})}}\dot{A}_{i}^{a}(\vec{x})\dot{A}_{j}^{b}(\vec{y}) \nonumber \\
    &&=\int d\vec{x}\; \Big\{2\langle \dot{W}^+,\dot{W}^- \rangle+\langle \dot{Z},\dot{Z} \rangle+\langle \dot{B},\dot{B} \rangle \nonumber \\
    && \qquad+ \langle A_0,[A_i,\dot{A}_i] \rangle+\langle A_0,J_0[h,h^\dagger] \rangle \Big\} \nonumber
\end{eqnarray}
One can then leverage (\ref{eq:a_0}) to filter through the $a=1,2$ gauge components of the connections in the last two terms above and hence obtain the non-flat contributions to the $W^\pm$ orbit space. Observe that the presence of the Higgs source involves time derivatives of $h$ and $h^\dagger$ and their mixture with $A_0$ will yield curved contributions to the Higgs field orbit space as well as warped geometry terms to the ($W^\pm,Z,B$) spaces. Consequently, the electroweak metric is not decomposable as a ``direct sum" of the metrics from each individual space
\begin{equation}
    \mathfrak{G}^{EW}\neq \mathfrak{G}^{W^\pm}\oplus\mathfrak{G}^{Z}\oplus\mathfrak{G}^{A}\oplus\mathfrak{G}^{h}
\end{equation}

Furthermore, the fact that the Higgs space is curved will mean that the Higgs boson will receive Ricci contributions to $m_h$ according to the gap theorem. The case is similar for the $W^\pm$ particles. Even if the source $J$ is ignored for ease, the nonflat terms of $\mathfrak{G}^{W^\pm}$ in the Coulomb chart are still horribly complicated functions of the fields ($W^\pm,Z,B$), their spatial derivatives, and the inverse gauge covariant Laplacian. Note that the decomposition of the metric into flat plus nonflat term would seem to suggest that the Coulomb gauge corresponds to a geodesic normal chart, hence curvature quantities would in principle be determined by explicit computation of the nonflat term. In particular, the task at hand would be to obtain the Ricci tensor and suitably regularize it by means of the point-split method. Once done, positivity would allow us to invoke the gap theorem and obtain the nontrivial intrinsic mass correction to $m_Z$. This in-depth analysis is left for a different time as the main purpose of this short address was to exhibit the ideas of how the mass anomaly might have a resolution via the curvature of the configuration space.

\begin{acknowledgments}
I thank P. Mondal from the Harvard Center of Mathematical Sciences and Applications for many fruitful conversations regarding his gap theorem and the arguments presented here. This project is supported by the Harvard College Research Program.
\end{acknowledgments}

% The \nocite command causes all entries in a bibliography to be printed out
% whether or not they are actually referenced in the text. This is appropriate
% for the sample file to show the different styles of references, but authors
% most likely will not want to use it.
\nocite{*}

\bibliography{apssamp}% Produces the bibliography via BibTeX.

\end{document}